# GobyWeb: simplified management and analysis of gene expression and DNA methylation sequencing data

Kevin C. Dorff[1], Nyasha Chambwe[2,3], Zachary Zeno[1], Rita Shaknovich[4], Fabien Campagne[1,2]*

[1]The HRH Prince Alwaleed Bin Talal Bin Abdulaziz Alsaud Institute for Computational Biomedicine. [2]Department of Physiology and Biophysics, [3]Tri-Institutional Training Program in Computational Biology and Medicine, [4]Department of Pathology and Department of Medicine;; The Weill Cornell Medical College, New York, NY, USA.

* To whom correspondence should be addressed (fac2003@campagnelab.org).

**Abstract.** We present GobyWeb, a web-based system to facilitate the management and analysis of high-throughput sequencing (HTS) projects. The software provides integrated support for a broad set of HTS analyses and offers a simple plugin extension mechanism. Analyses currently supported include quantification of gene expression for messenger and small RNA sequencing, estimation of DNA methylation (i.e., reduced bisulfite sequencing and whole genome methyl-seq), or the detection of pathogens in sequenced data. In contrast to many analysis pipelines developed for analysis of HTS data, GobyWeb requires significantly less storage space, runs analyses efficiently on a parallel grid, scales gracefully to process tens or hundreds of multi-gigabyte samples, yet can be used effectively by researchers who are comfortable using a web browser. GobyWeb can be obtained at http://gobyweb.campagnelab.org and is freely available for non-commercial use.

High-Throughput sequencing (HTS) instruments have been used to develop a variety of cost-effective assays. Each of these assays leverage the ability of second generation sequencing instruments to output millions of short sequence reads in a few days. As of the Fall 2012, it is not uncommon to generate about three billion 100 base pair long sequence reads per week with one HiSeq 2000 instrument (many core facilities have several similar instruments). Such throughput makes it possible to multiplex assays, which has contributed to reducing the cost of assaying each single sample. Reductions in sequencing costs are now making it possible for research groups to produce datasets with tens to hundreds of biological or clinical samples.

With increasing sequencing throughput, the management and analysis of large datasets produced with HTS assays have become a significant challenge for most research groups. Indeed, HTS data analysis is now recognized as a bottleneck of most research studies.

While many programs have been developed to process HTS data on the command line, only a few integrated systems have been developed that can help investigators process large amounts of data with a simple user interface. Existing systems with a user interface are often restricted to analysis of a single type of data (e.g., see [1, 2]), which forces users to work with different tools to analyze gene expression data or DNA methylation data, for instance. Systems that provide both a user interface and support multiple types of data have been offered commercially, but these systems often operate as black boxes and cannot be inspected in detail or extended.

We report the development of GobyWeb as a web application that can help users with no programming or command line experience analyze HTS datasets efficiently. GobyWeb takes advantage of compute grids to parallelize applications and dramatically accelerate computations for large datasets. This new tool provides intuitive and consistent analysis workflows that make it possible to track data and results for large projects. This report describes the user interface we have designed for GobyWeb, the types of analyses currently supported by the software, and the computational requirements for local installation.  We present examples of analyses that can be conducted with the system. A plugin mechanism is used to implement all types of analysis and makes it possible to customize or extend an installed instance of GobyWeb for future or custom analysis needs. Importantly, creating new plugins requires shell-scripting experience, but does not necessitate a strong parallel computing experience. We compare GobyWeb to several analysis software and systems previously described in the peer-review literature.

**Results**

**Software Overview.** We designed GobyWeb with the following main goals:
- Provide an intuitive user interface that biologists with limited bioinformatics experience can use effectively to analyze their datasets. Offer direct download of intermediary and final analysis results in well-defined formats to allow bioinformaticians to perform visualization or custom analyses.
- Support validated analyses for gene expression and DNA methylation.
- Provide mechanisms to track data. The system offers tags for each data element that can be recorded and used at a later time to retrieve data from the web interface. GobyWeb tags are listed in this manuscript following a description of an analysis and can be used to locate analyses in the GobyWeb demonstration system [3].
- Facilitate information sharing among members of a team of investigators.
- Offer efficient analyses that can process large datasets on small compute grids in a few hours.
- Support common types of analyses with scalable analysis workflows.
- Support future extensions via the definition of plugins for new alignment or analysis methods.
- Minimize the cost of data storage.

**User interface.** The GobyWeb user interface consists of the menus shown in Figure 1. The application menus are organized along three categories: Browse, Actions and Account. Figure 2 presents an overview of the data flows through the system and summarizes the types of interactions end-users have with the deployed system. We describe the options offered by these menus and data flows in Supplementary material as well as in video tutorials online (see http://gobyweb.campagnelab.org).

Importantly, this user interface is accessible from most modern web browsers. Upload of large data files must be done over a fast network connection at the start of an analysis. However, after uploads are completed, the system design makes it possible to perform analysis over connections typical of residential or mobile Internet access.

**System Architecture**. Figure 3 presents a high-level overview of the architecture of the GobyWeb software. Briefly, GobyWeb consists of a web front-end, a persistent data store, and a compute grid. The system uses production quality infrastructure components that are widely available in many academic institutions. The system can be installed within an institutional firewall, on an intranet or as an Internet facing application. Each user is required to obtain registration credentials with the system. User authentication enables rich data management capabilities and makes it possible to audit data and CPU usage on a per-user basis.

**Compute Grid.** We configured a GobyWeb instance with a compute grid of three nodes. Each node contains 4 Intel Xeon X5660 processors at 2.80GH and offers 24 effective threads and 48GB of memory (a configuration that was purchased for less than $9,000 per node three years ago, assuming a three year equipment life, the cost of each node is 34.22 cents per hour, excluding cost of electricity). This compute grid is referred in the following as the *small benchmark grid* and can run a total of 72 effective threads in parallel. It is difficult to ascertain the exact speed of the nodes used by published benchmarks performed in a cloud environment. For instance, in the Myrna benchmark evaluation [1], EC2 Extra Large High CPU Instances were used, but these 'instances' are virtual images and have been deployed on different hardware over the years. We are unable to provide a direct performance comparison with these tools because many of the cloud-based software cannot also be deployed on local servers. Nevertheless, for illustration purposes, we will compare the performance of the small benchmark grid to the 80 cores cluster mentioned in the Myrna benchmark (small cluster) [1].

**RNA-Seq data analysis.** As an illustration of the capabilities of the GobyWeb software, we uploaded 72 human mRNA-Seq samples, previously used to benchmark Myrna [1]. Reads were trimmed to 35bp and concatenated across the Yale and Argonne sites to closely replicate the benchmark conditions described by Langmead et al [1]. This dataset consists of approximately 1.1 billion 35 bp reads. Upload of these reads to GobyWeb took 30 minutes, compared to 1h 15 minutes as reported previously with Myrna [1]. We aligned reads to the genome with GobyWeb, using the BWA and GSNAP alignment plugins. The Myrna benchmark used the Bowtie aligner, which is often faster than BWA. Despite this difference, alignments with BWA and GobyWeb completed in 2h 22m, when Myrna reported 2h 56m (10 worker configuration). Detailed benchmark times are presented in Table 1, with data for Myrna obtained from [1]. Since we have performed our benchmark on different hardware than used in the Myrna benchmark, Table 1 is not an exact comparison but suggests that GobyWeb is competitive when aligning reads on small clusters. Because both BWA and Bowtie are unable to align RNA-Seq reads through splice junctions, we also aligned this dataset with GSNAP [4]. Spliced alignment with GSNAP was much slower, requiring 25h 20m to align the 1.1 billion reads, but as expected mapped many reads to the genome through splice junctions that BWA (and Bowtie) are unable to map.

The STAR aligner [5] is also available as a plugin to GobyWeb. STAR can perform spliced alignments, but in our configuration can only align reads longer than 50bp (because shorter reads are now uncommon). We used GobyWeb to align about 43 million reads with both GSNAP and STAR (publicly available dataset GEO GSM424349). Table 2 presents the duration of these alignments and indicates that STAR (tag: EBGNHJW) is about 4.8 times faster than GSNAP (tag: HJMAOVP) on this dataset, while providing comparable spliced alignments. To illustrate this later point, we visualized alignments generated with GobyWeb in the Goby format [6] with the Integrative Genomics Viewer (IGV, [7]). This comparison is shown in Figure 4.

Differential expression tests for genes can be conducted with GobyWeb using either DESeq [8] or EdgeR [9, 10]. A Goby differential expression plugin also makes it possible to estimate RPKM values and their logarithm for genes in individual samples and estimate fisher exact test statistics and non-moderated Student t tests (with adjustment for multiple testing with the Benjamini Hochberg method). While the Fisher statistic is not recommended for comparison of samples with biological variation [9], the RPKM values in individual samples are useful to create correlation plots. To assess the performance of different expression analysis for genes, we split the 72 samples in two groups (randomly, following [1]) and calculate differential expression with each method supported by GobyWeb. Table 2 summarizes these benchmarks. When Myrna performed the analysis in 80 minutes (10 node cluster), GobyWeb completed an equivalent analysis in 21 minutes (small benchmark grid).

**Pathogen detection.** Biological samples can be contaminated by viral or microorganisms other than the organism under study. When left undetected, such contaminations can bias the conclusion of a study [11]. Detecting pathogen contamination in clinical samples is also of great interest [12]. GobyWeb offers a plugin to detect pathogen contamination in samples. Briefly, alignments are processed to extract reads that did not align to the reference genome. Such reads are assembled and mapped to viral, bacterial or fungal transcriptomes. Results are summarized as a table of species matched by each sample or group of samples, contigs for assembled reads, and table of detailed contig mapping information. The process makes it possible to detect contamination by viral, bacterial or fungal organisms in a variety of samples. While several command line tools have been developed to detect pathogens in sequencing data (e.g., [11-13]), the pathogen detection plugin is tightly integrated with GobyWeb and makes it possible to routinely screen samples for pathogen contamination. To measure performance, we analyzed the 72 RNA-Seq Pickrell samples with the pathogen detection plugin (searching viral genomes, using the Minia assembler and stripping Illumina adapters from the reads prior to assembly). The analysis completed in 53 minutes on the small benchmark grid and detected two viruses and one phage (see Table 3). Detected viruses include the *Human herpesvirus 4*/ Epstein Barr Virus (EBV) with more than 10 contigs per sample, and the *Macacine herpesvirus 4*. Enterobacteria phage phiX174, sometimes used as spike-in for quality control on the Illumina platform was also detected in two samples. Detecting EBV in HapMap samples is expected because the HapMap cell lines were produced by transforming B lymphocytes with the EBV [14]. Detection of the *Macacine herpesvirus 4* (MH4) is likely to be artifactual, since MH4 is a virus of the same genus as the EBV, and is detected in each sample with less than 10 contigs.

**DNA Methylation.** GobyWeb supports the analysis of bisulfite-converted reads. DNA samples that have been processed with an experimental protocol such as RRBS, ERRBS or methyl-Seq make it possible to estimate methylation at specific cytosine bases in biological samples. DNA methylation analyses start with aligning the reads to a reference genome while allowing for the type of mismatches introduced during bisulfite conversion. To this end, GobyWeb offers a choice of alignment tools: GSNAP[4], Last [15], or Bismark aligner [16]. Each of these tools is implemented as a plugin, and it is easy to add support for new methods.

To measure the performance of bisulfite alignment with these tools, we uploaded 6 Reduced Representation Bisulfite Sequencing (RRBS) samples, the Dnmt samples [17], to GobyWeb. Each sample contained between 30 and 37 million reads. The six samples (201 million 36bp reads) uploaded in 23 minutes. Aligning these reads against the mouse MM9 genome required 14h and 3 minutes with GSNAP, 4h 13m with Bismark and 2h 33 minutes with LAST (see Table 4).

Analysis of DNA methylation data often requires estimating methylation rates at observed sites across the genome, and performing tests of differential methylation. GobyWeb offers two plugins to help with these analyses. The first plugin estimates methylation rates for each observed cytosine (base-level analysis). We simulated bisulfite conversion to compare methylation rates obtained with the Goby plugin to rates estimated from files produced with Bismark and found comparable agreement for both methods (Supplementary Figure 4). The second plugin estimates average methylation rates over a set of pre-defined annotations (e.g., CpG islands, promoter regions, gene body regions). Both these plugins perform tests of differential methylation when groups of samples are defined by the end-user. The plugins support up to 10 different groups and an arbitrary number of comparisons between pairs of groups (statistics of differential methylation are reported for each site for each comparison defined by the user). Results can be viewed and downloaded with web-based table views. GobyWeb table views are fully interactive, support filters on multiple columns, and scale gracefully to support results with hundreds of millions of rows (Figure 5). In addition to table view, both methylation plugins produce files suitable for visualization with IGV. The region-based methylation plugin produces files in the IGV format, while the base-level plugin produces VCF files that support viewing base-level methylation estimates across sets of samples. Figure 6 presents an example of visualization produced with these two plugins for the Dnmt samples.

Taken together, these capabilities are substantial improvements over software tools previously published.

**Discussion**

**Command line tools.** Development of HTS technology has spurred the development of specific computational approaches to process the data, such as alignment programs (e.g., BWA, Bowtie, or GSNAP) or approaches for calling differentially expressed genes (e.g., DeSeq, EdgeR). However, most tools were designed for users comfortable with the UNIX/Linux command line. This category of users rarely includes the biologists who generate the datasets. This fact contributes to creating an analysis bottleneck since bioinformaticians are needed even for the most routine data analyses. Beyond this mismatch, command line tools do not fully address the kind of practical problems that investigators encounter when their studies require the collection and analysis of tens of samples. This is especially true when each sample requires several Gigabytes of storage. For these projects, even experienced command line users can benefit from intuitive user interfaces that help with routine analyses, and can improve data organization and analysis reproducibility. In our experience, most researchers need data management capability to help with large HTS analysis projects. GobyWeb is an integrated analysis system that provides strong data management capabilities with a convenient user interface.

**Core Facility Pipelines.** Many bioinformatics groups have integrated command line tools into internal pipelines to facilitate data processing. Because pipelines are often implemented as scripts, the same limitations discussed for command line tools apply, and these pipelines are typically not exposed to biologists who generated the data, limiting results communicated to biologists to pre-determined sets of reports. In contrast to users of core facilities that maintain in-house pipelines, users of GobyWeb can query their own datasets directly, freeing time for bioinformatician to evaluate new methods, develop or install new plugins and generally focus on more interesting problems than running the same analysis ten times. The software also supports collaborative patterns where a set of users (e.g., members of a bioinformatics lab or core facility personnel) runs standard alignments and analyses for the type of data, and end-users browse and query the results in various ways, or run additional analyses trying different algorithms or parameters.

**Gene expression and DNA methylation.** The set of plugins distributed with GobyWeb provide state of the art methods for analysis of gene expression and DNA methylation data. STAR and GSNAP alignments can perform spliced alignments efficiently, DESeq or EdgeR statistics are available to call differentially expressed genes or splice sites with differential usage across groups of samples. Bisulfite converted reads can be mapped with Bismark, or the Last aligner, and analyzed across groups to yield differential methylation statistics at single cytosines or annotated regions. GobyWeb generates file formats that can be directly visualized in IGV to follow up on findings of differential methylation and integrate these observations with other annotations or data. Together these features provide an integrated analysis system to study gene expression and DNA methylation. We anticipate that methods developed as R scripts such as MethylKit or

**Cloud computing.** Several systems have been developed to run analysis of HTS data on compute clouds. Some of these systems provide capabilities to process collections of samples. However, these systems are often limited to one type of data and/or require users to transfer data beyond their institutional firewall. Example of such systems include Myrna [1], which focuses on RNA-Seq data, MethylKit, which focuses on base-level methylation data [18], SIMPLEX [2], which focuses on exome data, or Clovr, for bacterial genome assembly [19]. Systems that support a single type of data require users to work with multiple user interfaces for projects that require the integrated analysis of different assays. GobyWeb improves upon these systems by offering one convenient interface and numerous types of analyses. Our experience suggests that users can learn to use GobyWeb effectively in two, one hour, training sessions. Much of the material covered in these sessions is also offered as training videos online [3]. Cloud-based systems also often lack strong data management capabilities to help users work with many samples. A drawback of GobyWeb compared to cloud-based systems is that it currently cannot be easily deployed to a commercial cloud environment and is limited to a local grid. This is a drawback because cloud-based systems can procure on-demand compute capacity for periods when project activity spikes. We chose to focus on local grid deployment for the initial release of GobyWeb because the proximity of the analysis grid to sequencers deployed at an institution has significant performance advantages as the data volume of typical projects continues to grow. Costs of a local grid are also typically lower than cloud solutions when compute needs are sustained, and access to a server room and system administration team are available [20]. Because of its focus on internal grids, GobyWeb can be deployed in intranets with no Internet connectivity when data confidentiality is a strong requirement.

**Commercial systems.** Several proprietary systems provide data management features and are commercially available (e.g., GeneSpring NGS, Avadis NGS, Partek Genomics Suite). Beyond significant costs, commercially available systems are closed source, most rely on programs that were neither described nor evaluated in peer-reviewed publications, and many are not readily extensible to support new types of analyses. GobyWeb is offered free of charge for academic institutions, integrates many state of the art academic tools, and provides a mechanism to describe analyses as plugins, which are distributed with source code to facilitate peer-review and future extensions.

**Parallel computing.** We have designed the GobyWeb system from the ground up for parallel computing. A number of paradigms have been proposed to deploy HTS data analysis on parallel systems. Most command line tools developed by the bioinformatics community are either sequential or limited to node parallelism, where the program runs parts of the work on parallel threads on the same machine. This type of parallelism requires adding additional core or processors inside a single node to scale to larger datasets. In contrast, grid or cluster parallel computing paradigms can split workloads and coordinate a large set of nodes to achieve parallel speed-ups. GobyWeb takes advantage of the grid paradigm and makes it possible to reduce computing time by adding nodes to a compute grid. Several programming methods have been proposed to take advantage of collections of compute nodes. The MapReduce approach, described in [21] and applied to some bioinformatics problem efficiently co-locates data with compute resources. While very efficient, the approach requires most programs to be rewritten to fit the MapReduce requirements. In contrast to MapReduce, the GobyWeb plugin system can integrate and run in parallel a variety of software without major redesign and reimplementation of their algorithms. To achieve this, we require that the software is able to run on specific parts of very large files and to combine part results into a complete result set. This requirement is often much easier to meet than the typical algorithm redesign and re-implementation required for a MapReduce solution. A current limitation of the parallelization paradigm used in GobyWeb is that scripts must have access to a shared file system.

**Workflow systems** Workflow systems are used widely to interactively construct custom pipelines to integrate and/or query a variety of datasets with different tools. Taverna [22] and Galaxy [23] are well known workflow systems developed to support bioinformatics applications [22, 23]. Taverna relies on web services to perform computations. It is not well adapted to process large, multi-gigabyte, HTS datasets and can run into serious performance issues when the volume of data exceeds the capability of the network connectivity between the Taverna application and the services. In contrast, Galaxy can be installed locally to store large datasets and process them on a local compute grid. Galaxy provides a tool-box of several key HTS tools, which end users can apply to analyze their datasets. Because Galaxy is a general workflow system, it is inherently more flexible than GobyWeb, making it possible for end users to assemble custom analysis pipelines. In contrast GobyWeb is more rigid: it only supports a limited set of predefined analyses (plugins currently have strict sets of inputs and outputs). However, we have designed this limited set to cover a wide range of common HTS analysis, and have been able to optimize these analyses for very large HTS datasets. A key advantage of GobyWeb is that all data are stored in compressed binary formats [6], which are several orders of magnitude smaller than the uncompressed text files used internally by Galaxy (e.g., alignment files stored by GobyWeb are 10 times smaller than BAM files, which are themselves 3 times smaller than SAM text files or equivalent tab delimited files). Rigid GobyWeb type systems are complementary to general workflow systems: rigid systems can be rigorously tested, which is critical in some domains, such as clinical sequencing, while workflow systems make it possible to experiment with new analyses quickly. For instance, tab delimited data exported from GobyWeb after alignment and group comparison can be further analyzed in Galaxy with custom, project specific pipelines.

**Software Distribution**

GobyWeb can be downloaded from [3] under a non-commercial academic license. Detailed installation instructions are provided on the web site. GobyWeb plugins are distributed under the LGPL3 license and can be obtained from [https://github.com/CampagneLaboratory/gobyweb-plugins](https://github.com/CampagneLaboratory/gobyweb-plugins).

**Acknowledgements**

This investigation was supported by grant UL1 RR024996 (NIH/NCRR) of the Clinical and Translation Science Center at Weill Cornell Medical College, by grant LLS 6304-11 from the Lymphoma and Leukemia Society Translational Research Program, and by R01 MH086883 (NIH/NIMH). Support was also provided by the Tri-Institutional Training Program in Computational Biology and Medicine. We thank Dr. Ji-Eun Oh for suggestions for improving the GobyWeb user interface and Manuele Simi for help preparing the GobyWeb demonstration system.

**Table 1. Gene Expression Analysis performance.** Wall clock times for analysis of 72 Pickrell et al 36bp RNA-Seq samples.

|  | GobyWeb | | | Myrna | |
|---|---|---|---|---|---|
|  | System | BWA | GSNAP | System | Bowtie |
| **Number of nodes in benchmark system** | 3 | | | 10 | |
| **Number of cores in benchmark system** | 72 | | | 80 | |
| **Upload time** | 15m | | | 1h15m | |
| **Alignment time** | | 2h22m | 25h20m | | 2h56m |
| **Differential expression time** | 21m | | | 80m | |
| **Total time** | 36m | 142m | 1520m | 155m | 176m |
| **Analysis wall clock time** (including upload, alignment and differential expression tests) | | 225m | 1603m | | 331m |
| **Compute Cost per node** | | 0.3422 | 0.3422 | | |
| **Approximate cost for computation** | | $3.88 | $27.43 | $44.00 | |
| **Compute Infrastructure costs** (excluding storage, network and web) | $27,000 | | | $0 | |

**Table 2. Performance of spliced alignments with GNSAP and STAR.** Alignments were performed with the GobyWeb and the GSNAP or STAR alignment plugin. One 50bp single end RNA-Seq sample with about 43 million reads.

| Aligner | GSNAP | STAR |
|---|---|---|
| **Wall-clock time** for alignment and statistics collection | 373m | 78m |

**Table 3. Pathogen detection performance.** Pathogen detection took 53m for the 72 Pickrell et al RNA-Seq samples. N: Number of samples where GobyWeb identified at least one viral contig from the specied organism. See tag: DNOAOZI.

| Viral organism detected | N | Comments |
|---|---|---|
| Human herpesvirus 4 type 1 (EBV) | 72 | The HapMap lymphocyte samples were transformed with EBV to yield individual lymphoblastoid cell lines. Detecting EBV in these samples is therefore expected. |
| Human herpesvirus 4 (EBV) | 72 | |
| Macacine herpesvirus 4 | 23 | Likely mis-detected because of close homology with the EBV virus (less than 10 viral contigs per sample are detected in a subset of samples). |
| Enterobacteria phage phi X 174 | 2 | Likely spike-in with Ilumina PhiX phage DNA. |

**Table 4. DNA methylation analyses.** We analyzed 6 RRBS samples organized in two groups to detect differentially methylated regions and bases.

| | System | GSNAP | Bismark | Last | GobyWeb Analysis of individual cytosines | Analysis of annotated regions |
|---|---|---|---|---|---|---|
| **Upload** | 78m | | | | | |
| **Alignment** | | 14h03 | 4h13 | 2h33 | | |
| **Differential Methylation Analysis** | | | | | 17m | 5m |
| **Complete base level analysis, wall clock time (minutes)** | | 860 | 975 | 170 | | |

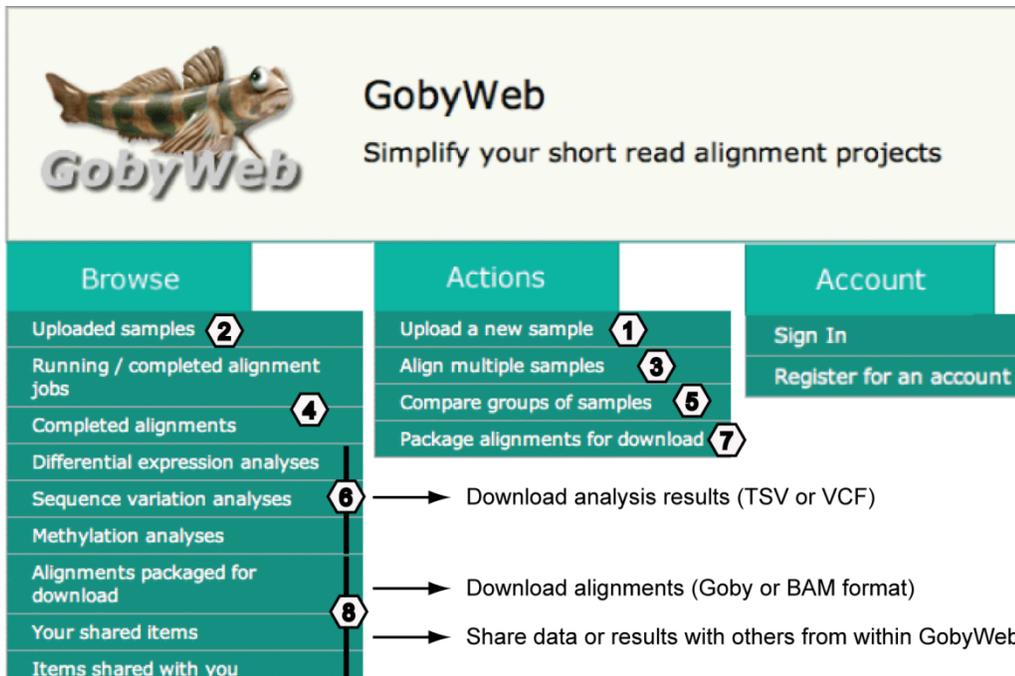

**Figure 1. GobyWeb user interface menus.** Increasing numbers indicate the typical order in which a user would navigate the interface, from data upload (1) to download or sharing results with others (8). See Supplementary Material for a detailed description of each step.

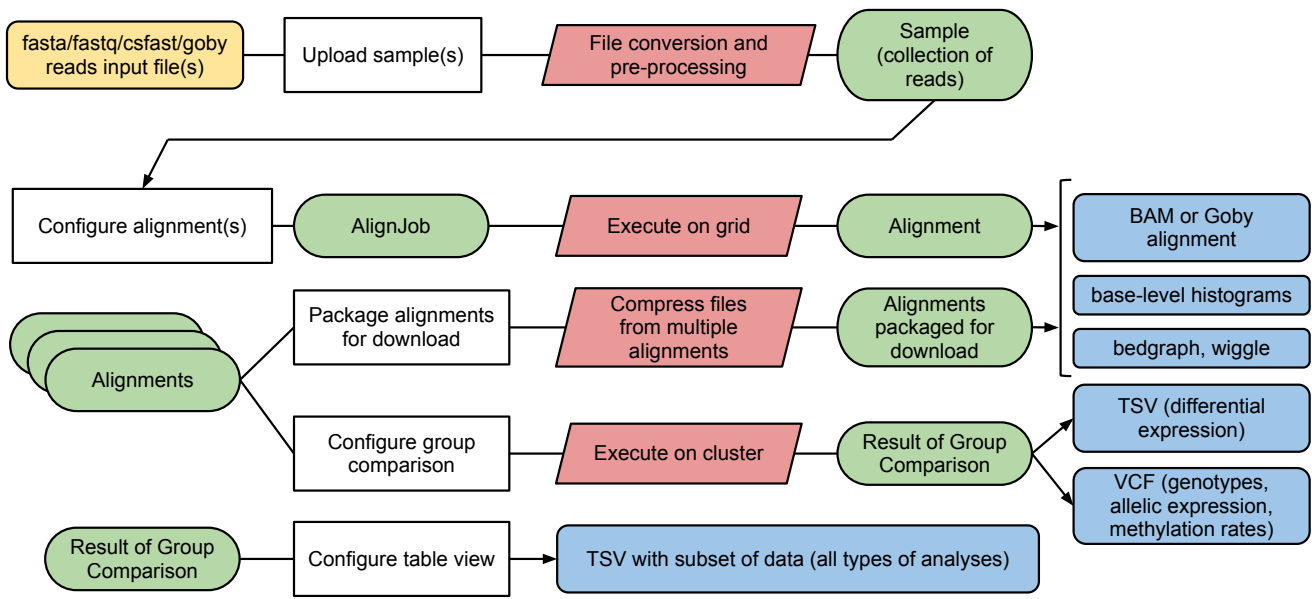

**Figure 2. Overview of flows of data in the GobyWeb system.** A typical project starts with upload of data files (in yellow, top left). Tasks that run on the compute grid are shown in red. Items of data represented in GobyWeb are shown in green and have dedicated web user interface views. Most views offer the option of downloading result files (in blue) in formats compatible with third-party software.

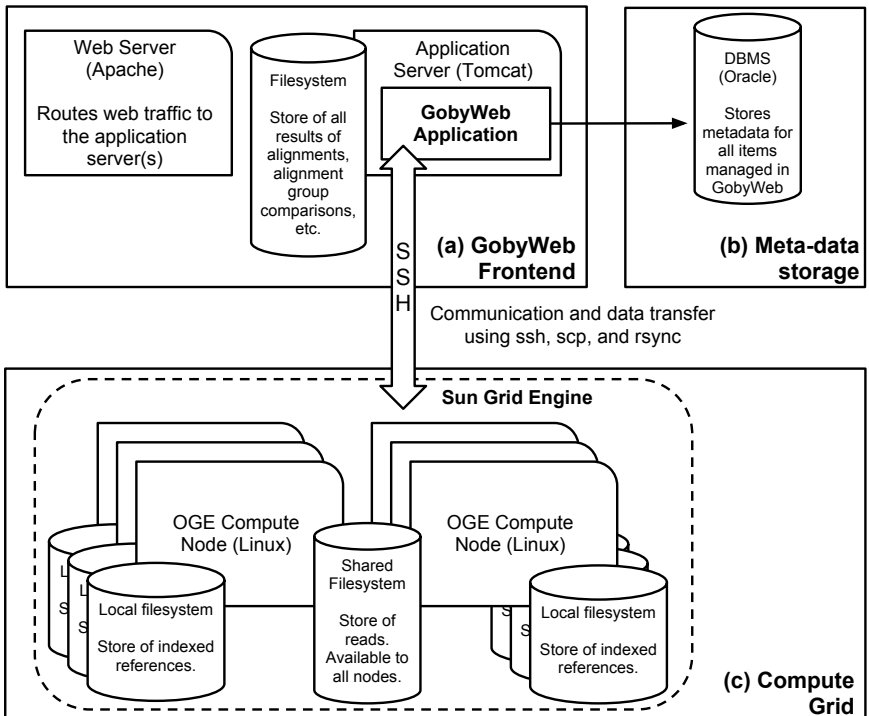

**Figure 3. Overview of the system architecture.** An installation of GobyWeb relies on three pieces of infrastructure. (a) The web front-end is deployed as a Java web application on one or more application server(s). Several servers can be used to scale the application up under heavy usage. (b) Meta-data about samples, alignments, analyses and users are stored persistently in a Database Management System (DBMS). (c) A compute grid is used to process large datasets efficiently. All datasets (reads, alignments, processed results) are stored as large files on local disks directly attached to each compute node, and the web application servers, as well as in a shared network file system. The software automatically performs data transfers between the shared file system and local storage disks and optimizes these transfers to maximize the overall analysis throughput of the system. The system relies on production quality software components (Apache web server, Tomcat application server, Oracle/JDBC DBMS, and Sun Grid Engine, Linux and Network File System) that are already available and used in many academic institutions.

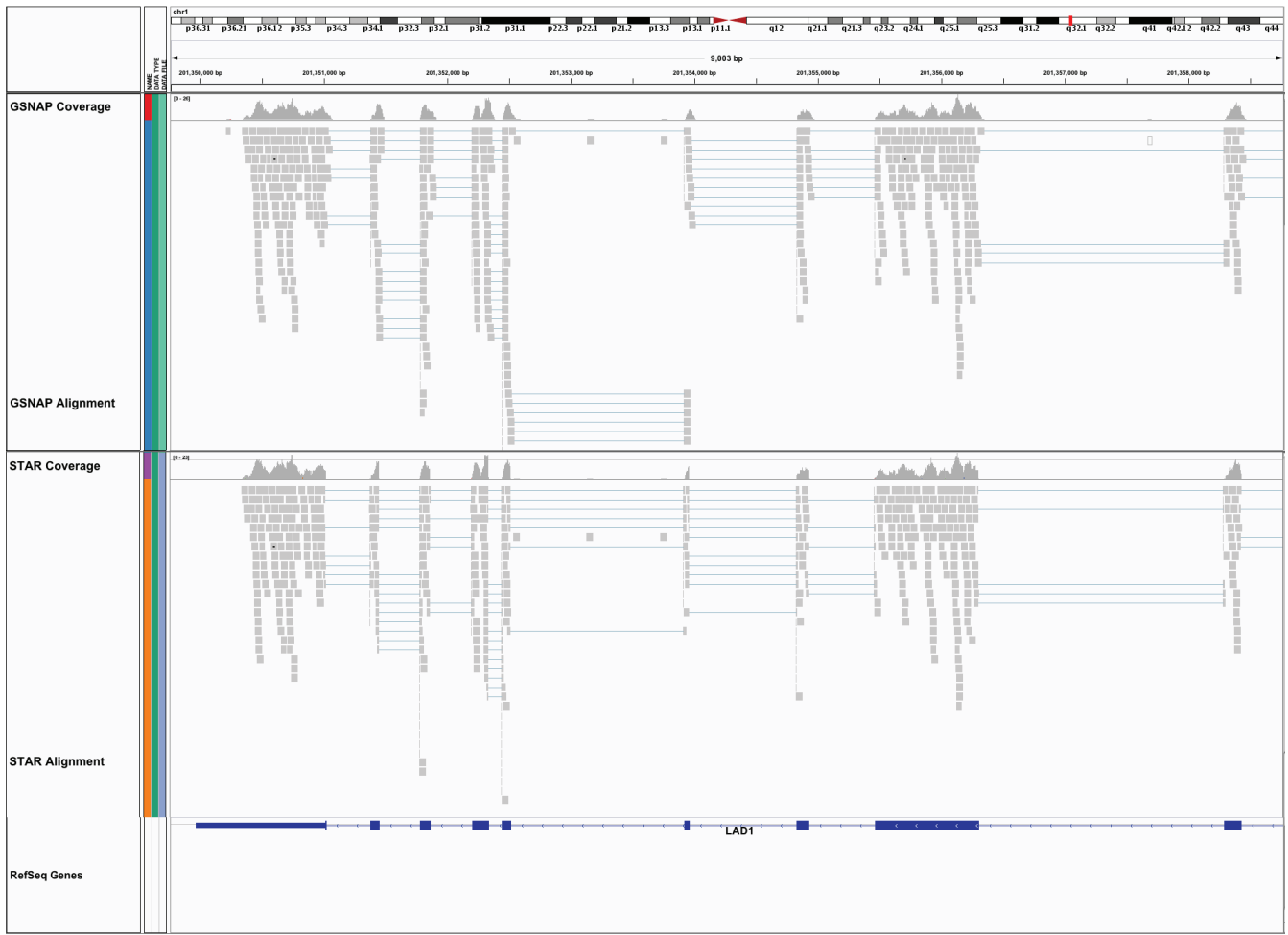

**Figure 4. Comparing RNA-Seq alignments done with GobyWeb and the GSNAP and STAR aligners.** This figure was constructed with the Integrative Genomics Viewer (IGV), which directly supports alignments in Goby format. Alignments in the Goby format are substantially smaller than in BAM format, and can be directly downloaded from GobyWeb for interactive visualization with IGV. The plot compares spliced alignments generated with the GobyWeb GSNAP and STAR plugins over the LAD1 gene.

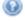

**Figure 5. Scalable Table Views.** GobyWeb offers web-based table views that scale to support tables of results with hundred of millions of rows. Users can subset the table to keep specific columns, as well as rows that match complex filters on column values. This mechanism makes it possible for end-users to work with very large tables and download only interesting subsets of the data, even over slow Internet connections. In this snapshot, the table viewer displays results from a base-level methylation analysis (tag=RQLDONK). The panel "Filtered list of elements" displays the current view of the table. The panel at the bottom makes it possible for end-users to select which subset of columns they need to visualize/download. The filters help users identify columns by keyword. Text boxes under each column are used to enter filtering criteria on the specific column.

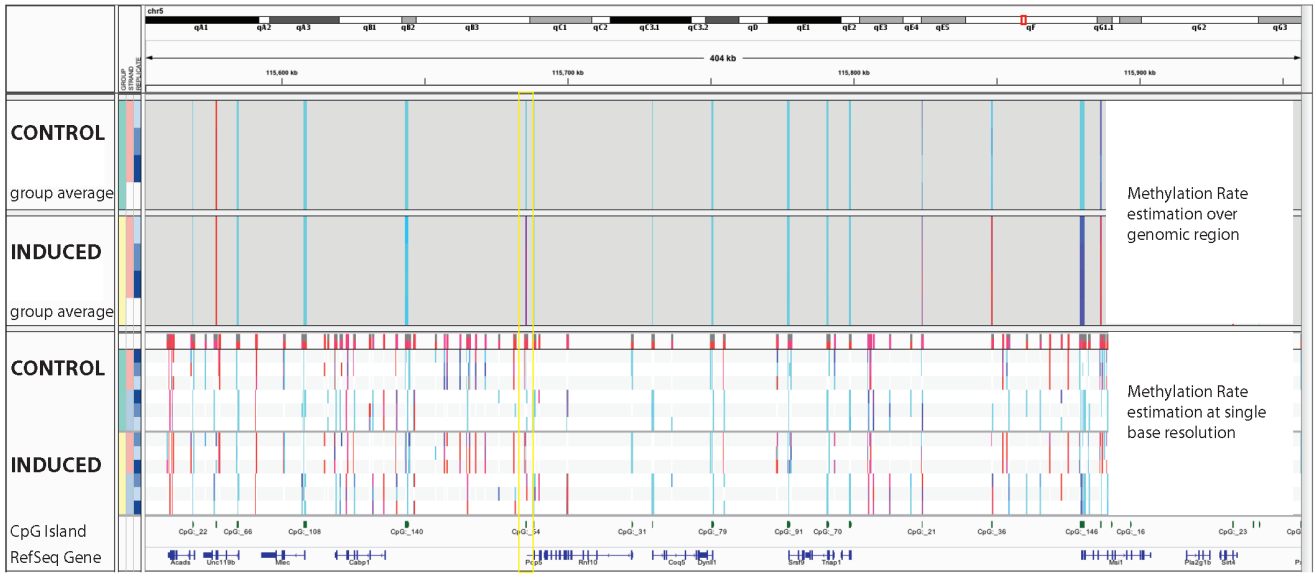
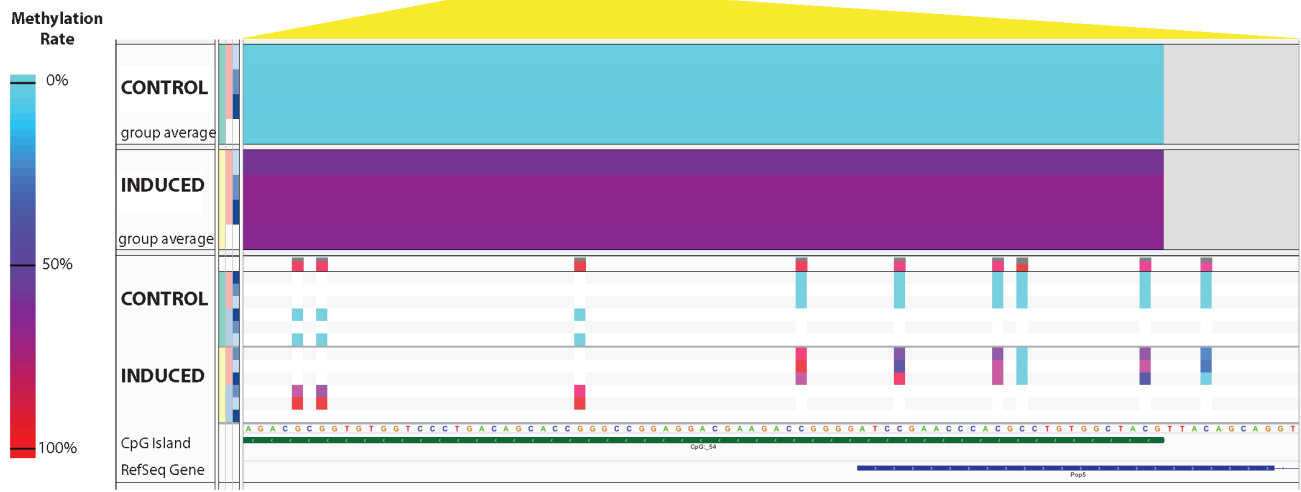

**Figure 6. DNA methylation data analyzed with GobyWeb and visualized with IGV.** GobyWeb produces data files in formats directly supported by the Integrative Genomics Viewer (IGV). This figure presents the results of methylation analysis over regions and individual bases for the Dnmt public datasets [17]. The bottom insert shows a smaller region with more details of the methylation rate at individual bases. Three rows per strand are shown, corresponding to 3 control and 3 induced samples. Integration with IGV makes it possible to visualize DNA methylation rates alongside other types of annotations or data types supported by IGV. The genomic region shown was selected among the regions that show one of the smaller p-values when comparing the control and induced group (empirical p-value, GobyWeb).

Supplementary Material for:

# GobyWeb: simplified management and analysis of gene expression and DNA methylation sequencing data

Kevin C. Dorff[1], Nyasha Chambwe[2,3], Zachary Zeno[1], Rita Shaknovich[4], Fabien Campagne[1,2]*

[1]The HRH Prince Alwaleed Bin Talal Bin Abdulaziz Alsaud Institute for Computational Biomedicine.[2]Department of Physiology and Biophysics, [3]Tri-Institutional Training Program in Computational Biology and Medicine, [4]Department of Pathology and Department of Medicine; The Weill Cornell Medical College, New York, NY, USA.

* To whom correspondence should be addressed.

## Software functionality

In the following description, we will refer to biologists who use GobyWeb for data analysis as end-users. We will also refer to users who install and maintain an instance of GobyWeb on a local computing infrastructure as administrators.

**Typical session.** During a typical analysis project, a GobyWeb end-user often performs the following steps:
1. Imports data into GobyWeb (Actions->Upload menu)
2. Inspects uploaded data (pre-alignment quality control)
3. Aligns samples against a reference genome
4. Inspects alignment results (post-alignment quality control)
5. Compares groups of samples (the type of analysis varies with the kind of samples being compared)
6. View analysis results in web-browser or download results for processing with other tools
7. Package alignments for archival or to process alignments with custom scripts on a local machine
8. Share results or data with other users of GobyWeb, granting or obtaining access.

**User registration.** The Account menu makes it possible for new users to register with an instance of GobyWeb. Registering requires an invitation code, a username, a password and an email. The invitation code restricts registrations to those users who are invited to use the system. It is possible to disable the invitation code or print it on the registration page to entirely open the system. Registration makes it possible for the application to associate datasets to an owner, and formalize the concept of data sharing. Emails are collected to make it possible for the system to send notifications to end-users when analysis jobs are started or completed. Collecting emails also helps administrators notify end users in preparation of periodic system maintenance.

**Uploading data.** HTS reads can be uploaded to GobyWeb directly from the web browser (menu 1 in main manuscript Figure 1). The upload page uses a Java applet to make it possible to upload multi-gigabyte files. Files can be uploaded in compressed fasta, fastq or csfasta, fastq.gz.tar (produced by the Illumina CASAVA 1.8 pipeline) or in the Goby compact reads format. As an alternative to uploading via the web browser, GobyWeb supports two methods for uploading datasets from the command line. The first method is designed for sequencing facility staff members to transfer datasets directly to a GobyWeb user account. The second method makes it possible for individual users to upload data to their account from the command line, and is useful when end-users access the user interface via a network connection that cannot transfer large amounts of data (e.g., using a home network to upload and start an analysis when raw data resides within the institutional firewall).

**Meta-data.** The upload page makes it possible to describe meta-data about the read datasets. This option is available irrespective of the manner in which the data were uploaded. The following meta-data can be described: the technology used to sequence the reads (Illumina, SOLID, Helicos, Roche), the species/organism and the tissue whose biological material was assayed, a textual description of the protocol of the experiment, as well as whether the protocol used for library preparation preserved the strand. Some of these options will affect how GobyWeb runs alignments or conduct analyses and therefore should match the nature of the data uploaded.

**Samples.** GobyWeb defines the concept of a sample as the set of reads sequenced from the same biological or clinical material, for the purpose of analysis each sample should be considered homogeneous and independent of other samples. The check-box at the top of the upload page triggers the creation of independent samples when multiple files are uploaded, or the concatenation of the reads into a single sample. Samples whose reads are derived from DNA that has been bisulfite treated (i.e., Methyl-Seq or RRBS) can be marked as such. When files with the extension .fastq.gz.tar are uploaded, these files are assumed to have been produced by the CASVA 1.8 pipeline. In this case, the reads in each tar file (set of fastq.gz files) are concatenated to produce one sample. Sample upload also accepts a tab separated value file (.tsv extension) with a set of samples that can describe more custom read pairing and concatenation needs. The format of the TSV file is described online.

**Data Sharing.** Users can indicate that other end-users be given access to a dataset they own. End-users can share data at upload time, or at any time after upload by editing the dataset and adjusting the list of users that the item is shared with. Only owners of a data item can adjust sharing for this data item. When user A shares data with user B, the data appears to B as owned by user A. Clicking on a username in a list of data items provides information about the owner, including an email address. Users of GobyWeb can therefore communicate by email to request that another set of users be given access to a dataset by the owner.

**Pre-alignment quality control.** After samples are created, GobyWeb will start background analysis tasks that inspect the read files and collect quality control statistics. While this process is ongoing, the samples appear to the end user in the list of uploaded samples, but are marked as "not ready to align". After the process completes, samples appear "ready to align" and can be selected to view collected statistics, or start alignments.

**Aligning reads.** End-users can align samples against a reference sequence. Users can either align a single sample by clicking on the Align link at the bottom of the sample page, or select the Align multiple samples option in the Actions menu. The alignment configuration page is shown in Figure 2. End-users can select one of several aligners: bwa (recommended for DNA-Seq), STAR or GSNAP (recommended for mRNA-Seq, GSNAP also supports bisulfite treated reads), last (recommended for smallRNA-Seq or when aligning to a reference different than that of the reads, also supports bisulte treated reads), or Bismark (for bisulfite converted reads). The subset of samples matching a specified organism is shown in a user interface component that supports interactive pattern matching (the list of samples is updated after every key-stroke with the set of samples that match the typed filter).

**Alignments.** Upon saving an alignment job, GobyWeb schedules the execution of the alignment in parallel on a compute grid. The end-user can obtain the status of submitted jobs in the "Browse>Running / completed alignments" menu item. Each individual alignment job also provides the status of its parallel computation on the grid (Figure 3). Upon completion, alignments can be downloaded in the Goby format, or in BAM format, either individually, or packaged as a ZIP file (archives can be produced with the "Action>Prepare alignments for download" menu item). Alignments are also post-processed to yield base level histograms in the Goby Count format.

**Group comparisons.** GobyWeb provides a multi-step software wizard to help end-users configure analyses that compare groups of samples. The wizards proceed in the following steps:
1. Define information about the group comparison analysis (e.g., number of groups to be compared, organism of the samples to be compared, description of the comparison analysis, type of analysis to perform, output format).
2. Select samples for the first group and name the group, repeat step 2 for each group defined in step 1.

3. Specify which pairs of groups need to be compared.  Zero or more comparison pairs can be specified at this step. For instance, when comparing four groups (A, B, C, D), three groups may need to be compared to a reference group (i.e., A/B, A/C, A/D), or groups may be compared two by two such as (A/B, B/C, C/D).
4. Review the association between groups and samples, the pairs under comparison, and submit the analysis for execution.
5. Review results in table view, filter and download subset of tables, alternatively, download entire table in tab delimited or Variant Call format.

The following types of analyses are currently supported:
- Differential expression for gene, exon or CNV regions. Count of reads that overlap a region are compared across groups and statistics of differential expression evaluated.
- Differences in allelic expression (RNA-Seq). Heterozygote sites can be tested for difference in allelic ratios between groups. This analysis identifies heterozygous sites where the proportion of reference allele differs significantly between groups.
- Methylation rate analysis for individual bases or annotated regions. Alignments from bisulfite-converted samples can be used to estimate methylation rates at the cytosines sites of a genome and identify cytosine bases whose methylation rate differ significantly across groups.
- Calling genotypes. Genotype calls at all sites with at least 10x coverage can be generated. TSV and VCF files can be generated.
- Calling sites where alleles associate with groups. Genomic sites are identified where the count of alleles (expressed in number of samples with the allele in the group) differ across groups.

**Exportable file formats.** Intermediary and final results can be exported from GobyWeb to facilitate visualization or enable additional custom analyses.

*Alignments* and derived information can be packaged as Zip files and downloaded (point 7-8 in main manuscript Figure 1). Alignment can be downloaded either in the Goby format, or in BAM format (when the alignment was generated with the BAM output options shown in Figure 2). Both formats can be visualized with IGV to inspect specific regions of the genome and view mapping of individual reads.

*Histograms.* Goby base level histograms store the number of reads that cover a reference sequence, at every base of the reference, and can be downloaded and viewed as coverage tracks with IGV. Histograms are also produced in Wiggle or bedgraph format for visualization on the UCSC genome browser, or to process with bed-tools [1]. The Goby histogram format is recommended for viewing coverage in IGV because such files are typically 5 times smaller than equivalent bedgraph files.

*Variant Calling Format.* Genotypes, results of comparison of allele-specific expression, or differences in methylation rates are exportable either as tab-delimited files (from the GobyWeb table viewer) or in the Variant Calling Format (VCF 4.1) [2]. VCF files produced by GobyWeb are annotated with respect to which gene overlaps the genomic site described, the RefSNP (rs) number of the variation, and the expected effect of the variation (data obtained when available from Ensembl and biomart for the human reference genomes). VCF files are indexed and end-users can download both the VCF file and its index. VCF files produced by GobyWeb are compatible with IGV. An IGV extension recognizes the methylation rate format produced by GobyWeb and renders methylation rates as color gradients in a VCF track.

## Methods and Software Implementation

**Web application.** GobyWeb is implemented as a web application written with the Grails framework (http://grails.org/). We have deployed the application in a Tomcat application server. Object persistence was implemented with Hibernate (http://www.hibernate.org/) and an Oracle backend. HTS read files are multi-gigabyte files that frequently exceed the 2GB size limit that HTTP file uploads can support. Large file uploads are supported in GobyWeb with the Jumploader applet (http://jumploader.com/).

**User interface design.** The application was developed iteratively and the user interfaces were designed as a team, then implemented and continually adjusted in response to user feedback. Initial releases were limited to the Campagne laboratory and direct collaborators, but a version of GobyWeb has been released to a wider audience of more than 70 investigators (faculty, post-doctoral fellows and students from the Weill Medical College of Cornell University, Memorial Sloan Kettering Cancer institute and Hospital for Special Surgery) since January 2010.

**Grid computing.** Embarrassingly parallel analyses are split into chunks and scheduled as array job on an Sun/Oracle Grid Engine (OGE, http://www.oracle.com/us/products/tools/oracle-grid-engine-075549.html). OGE supports arbitrary job dependencies and this feature is used extensively to interleave user jobs and increase overall job throughput on the compute grid. Results of computations for independent chunks are merged by a post-process OGE job that is configured to start when all parts of the array job have finished. Importantly, the post-process job is configured to start even if some parts of the array job fail. This is useful when working with aligners or other software that sometimes can reproducibly fail when run on specific data items (in such cases, GobyWeb produces an output for the data that could be processed, and provide visual indication of failure for the chunks of the parallel job that failed to compute). We use the Goby framework (http://goby.campagnelab.org, [3]) to split read and alignment files and combine results efficiently.

**Computational analyses.** Logic for computational analyses is implemented in Bash scripts. These shell scripts automate data transfers between the web server and the compute cluster as required by end-user analyses. The scripts schedule long running and parallel jobs on the Oracle Grid Engine instance with appropriate dependencies for each type of analysis. Scripts are also responsible for informing the web application of changes in the analysis status (job submission, start, intermediate steps, failure or completion). GobyWeb bash scripts are organized with a plugin architecture that makes it possible for developers and administrators to extend GobyWeb with custom alignment or analysis methods.

**Plugin architecture.** GobyWeb currently supports three types of plugins: resource, aligner and alignment analysis. Plugins are organized in directories that contain a plugin definition file (config.xml), a plugin script (script.sh) and optional plugin files (named as described in the plugin definition file). Briefly, resource plugins provide access to data files or executables that other plugins can share access to, such as samtools, vcf-tools or executables for alignment programs. Aligner plugins make it possible to extend GobyWeb to perform alignments with new alignment tools. Aligner plugins can generate either BAM or Goby alignments and run each sample file either sequentially, or in parallel. Alignment analysis plugins make it possible to extend GobyWeb with methods for analysis a set of alignment results. Source code for GobyWeb plugins is distributed at GitHub (https://github.com/CampagneLaboratory/gobyweb-plugins). Developing new plugins consist in creating an XML plugin configuration file and providing a bash script with predefined shell functions that interface between GobyWeb and the analysis or alignment tool.

**Plugins and parallelization.** Both aligner and alignment analysis plugins can be written to run either in a node-parallel mode, or in a grid-parallel mode. The node parallel plugins can take advantage of thread parallelism on a single node. They are simple to implement because it is only necessary to define one process function per plugin. However, node parallelism limits the maximum data processing throughput that can be obtained when a grid with several nodes is available. Node parallel plugins are also only efficient if the programs they use can take advantage of thread parallelism. In order to support programs that run sequentially, and increase data processing throughput by using more nodes of a compute grid, the plugin system also supports grid-parallel plugins. Grid-parallel plugins need to implement four shell functions, which are used to (i) determine how to split a files into chunks, (ii) count the number of chunks generated in the first step, (iii) process a chunk to produce a result, and (iv) combine many results to produce a complete result.

**Status monitoring.** Status updates are communicated to the web application by posting messages to pre-defined URLs. Scripts and plugins accomplish this by calling a command line tool with status information and the tag of the analysis that the script is running. The web application stores status updates in the Hibernate persistence store (e.g., oracle database). Upon receiving a completed status update, the web application checks that result files have been created and are valid. When this is verified, the information about the job is saved to the Hibernate persistence store.

**Read storage.** Upon sample upload, fasta, fastq or csfasta read files are converted to Goby compact-reads format with the fasta-to-compact tool [3]. This transformation is performed to allow efficient parallelization of the alignment jobs (see Grid computing and Read alignments sections). Reads are transferred from the web application server to the shared file system visible to the compute cluster (copies are performed with scp). A number of pre-processing steps are executed to validate input files, obtain read length and base quality statistics, as well as associate weights to reads (such as heptamer weights required by the Hansen et al approach [4]). Read conversion and processing steps are executed on the OGE grid and proceeds in parallel for each sample file uploaded.

**Read alignments.** For plugin aligners that support parallel processing, alignments are split as OGE array jobs with *n* tasks, and a post-process job (see Grid computing section). The number of tasks is determined by dividing the read file size by the chunk size (in practice, we use CHUNK_SIZE=50,000,000, for approximately 50MB chunks). When an array job component starts executing on the cluster, OGE sets the SGE_TASK_ID variable to the index of the task in the array job. We determine the start and end offsets within the reads file that contains the reads that this task should align as follows:
START_POSITION=(SGE_TASK_ID - 1) * CHUNK_SIZE. Similarly, END_POSITION= START_POSITION + CHUNK_SIZE - 1. We then run the aligner plugin 'plugin_align' function with these START_POSITION and END_POSITION values, since these options restrict the alignment to only the reads contained within these file byte offsets. When the array tasks finish, Goby alignment results are concatenated with the concatenate-alignment Goby tool, calling the tool recursively to combine at most 100 pieces at a time and therefore limiting the number of files open at any given time. This process makes it possible to scale alignment concatenation to alignments with billions of reads. To support aligners that do not provide native Goby support, we use the goby compact-to-fasta tool with –s START_POSITION and –e END_POSITION to extract Fasta or Fastq format for the reads of the part. Aligners that do not produce Goby alignment format require the implementation of conversion scripts that convert the aligner output to Goby format. An example of this method is provided for the last/lastag aligners. For plugin aligners whose definition file indicates that they do not support parallel processing, we run the alignment of each sample as an independent OGE job with exclusive access to a node, and call the plugin_align function of the plugin with the entire reads file. Such aligner plugins are expected to leverage multi-core architectures on each node.

**GSNAP plugins.** The GSNAP aligner is integrated with GobyWeb in the GSNAP_GOBY and GSNAP_BAM plugins. The former generates alignments in the Goby format, while the latter yields alignments in the BAM format. Both plugins require a recent version of GSNAP compiled with Goby support. The GSNAP aligner, or the STAR aligner, are the recommended choice when aligning RNA-Seq samples to a genome because they can perform spliced alignments efficiently (i.e., map reads that span one or several exon-exon junctions). GSNAP also supports mapping reads from samples converted with sodium bisulfite (e.g., as in the Methyl-Seq and RRBS protocols). When aligning bisulfite converted samples, the GSNAP_GOBY plugin aggressively trims reads to remove adapter sequences (these sequences have to be provided by Goby administrators and can be obtained from the vendor of the sequencing instrument and kits). Adapter trimming is performed with the Goby trim mode and only trims adapters if the trimmed part of the read would exceed four base pairs. Trimmed reads are then aligned with GSNAP in methylation mode with parameters that disable indel matches and terminal trimming (--mode cmet -m 1 -i 100 --terminal-threshold=100).

**BWA plugins.** The BWA aligner is integrated in the BWA_GOBY and BWA_BAM plugins. These plugins require the version of BWA patched to support Goby file format (available from http://campagnelab.org/software/goby/bwa/). BWA_GOBY is grid-parallel and produces Goby alignment files, wile BWA_BAM is node parallel and produces BAM alignment files. BWA plugins are recommended when aligning DNA-Seq samples.

**Last plugin.** The LAST_GOBY plugin runs a recent version of the Last aligner (currently version 230). This aligner is recommended when the reads are from an organism for which no reference genome is available and when one needs to align to a reference that is a distant homolog from the organism of interest. The Last aligner is also recommended when aligning small RNA sequences.

**Last bisulfite plugin.** The LAST_BISULFITE plugin runs the Last aligner with matrices suitable to align bisulfite converted reads. The plugin searches both forward and reverse strands of a specially indexed reference sequence, and combines the results following the recipe described at http://last.cbrc.jp/doc/bisulfite.txt.

**RNA-Seq.** GobyWeb currently offers five plugins to analyze RNA-Seq results. These plugins calculate counts in parallel, combine results from parallel splits, normalize counts and estimate statistics of differential expression. DIFF_EXP_GOBY estimates counts over annotations with the Goby alignment-to-annotation-counts. This plugin outputs counts, RPKM and $\log_2$(RPKM) for each alignment included in the analysis. This plugin also estimates Student t-test statistics for RPKM values between two groups and fisher-exact test on the raw counts. DIFF_EXP_DESEQ estimates counts over annotations with Goby, but uses the R package DESeq to estimate statistics of differential expression. The third plugin DIFF_EXP_EDGE_R integrates the EdgeR package with GobyWeb (count estimation is performed with Goby). The fourth plugin, SEQ_VAR_GOBY has an output format that estimates allelic differential expression. In RNA-Seq data, this plugin estimates the significance that the reference allele expression is different between the groups under study. The plugin is implemented with the Goby discover-sequence-variant mode –format allelic-frequencies. The fifth plugin, SPLICING_DIFF_EXP, counts the number of reads spanning splice junctions to determine alternative splicing usage (counts spanning junctions are determined with Goby and statistics of differential splicing usage are determined either with DESeq or EdgeR).

**RNA-Seq normalization.** GobyWeb supports four normalization methods: hexamer bias removal[4], RPKM/FPKM normalization, upper-quartile normalization[5] and the TMM procedure implemented in EdgeR [6, 7]. The method of Hansen et al [4] was implemented to make it possible to remove random hexamer priming biases. Heptamer weights are associated to each read during the post-upload process are used as described previously ([4]) to estimate bias adjusted counts. RPKM normalization is conducted with the Goby alignment-to-annotation-counts and calculated as $r=[(c+1) / ( L / 1000.0) / (N / 1.10^6)]$, where c is the count, or number of read fragment that overlap with a given annotation, L is the length of the annotation and $N$ a normalization factor. For FPKM/RPKM normalization, N is taken to be the total number of read fragments aligned in a sample. The upper-quartile normalization method is implemented with the same formula, but using the 75 percentile of annotation counts as the value of N, as described previously by Dudoit and colleagues[5].

**Pathogen detection.** To determine the presence of pathogen in sequenced samples, this plugin (CONTAMINANT_EXTRACT) proceeds in three steps. (1) Reads that do not map the reference sequence (unmapped reads) are retrieved from a set of alignments and reads files. (2) Unmapped reads are optionally trimmed from adapters. (3) Unmapped reads are assembled into contigs. GobyWeb can use either the Trinity assembler or the Minia assembler[8]. Minia is the default choice and recommended for performance and memory usage. (4) Contigs are used to search a large and diverse database of (a) viral, (b) bacterial, or (c) fungal organisms. In the current version of GobyWeb, viral, bacterial and fungal RNA sequences are used and obtained from RefSeq [9]. Contigs that match in these databases are considered annotated if they match over more than 150bp and have an E-value less than 1e-6. The species matched by annotated contigs in these diverse databases are recorded and associated to the sample that contributed the unmapped reads. Contig sequences are provided in Fasta format.

**DNA-Seq.** GobyWeb offers a few approaches to analyze DNA-Seq data. A samtools plugin (SEQ_VAR_SAMTOOLS) makes it possible to call genotypes with samtools mpileup, or to estimate allelic association tests for alignments in the BAM format. The SEQVAR_GOBY plugin supports calling genotypes as well as performing allelic association tests for alignments in the Goby format. The SEQVAR_GOBY plugin is grid-parallel and optimized for comparisons involving large numbers (>50) of samples across groups. **This plugin should be considered experimental until we publish the results of large-scale validation tests.** The output is generated for both plugins in the Variant Calling Format 4.1 (VCF).

**Methyl-Seq.** GobyWeb estimates methylation rates and calls sites of differential methylation across groups of samples. These analyses are implemented in the SEQ_VAR_GOBY_METHYLATION plugin. This plugin uses the Goby discover-sequence-variants mode with the methylation output format. This mode implements statistics of differential methylation with a fisher exact-test at individual genomic positions where at least one cytosine is observed. The use of fisher exact test statistics to call differential methylation has been described recently [10]. P-values are adjusted for multiple testing with the Benjamini-Hochberg method across all sites tested in the genome. Methylation rates and statistics are written to VCF format. This plugin also support an empirical p-value estimation (empirical-p) methods that takes into account biological variability within groups and will be described elsewhere.

**Variant Calling Format.** VCF files are annotated with VCF-annotate to map sites of variations to gene, RefSNP rs id identifiers and variation predicted effect, when possible. Annotated VCF files are then sorted in genomic position (vcf-sort) and indexed with tabix [11]. This process makes it possible to load VCF files produced with GobyWeb directly in the Integrated Genome Viewer (IGV). We have extended IGV to display VCF files that encode methylation rates, as produced by GobyWeb and Goby.

**Other plugins.** We frequently add new GobyWeb plugins or improve existing ones. The definitive source of information about plugins is the GitHub repository at https://github.com/CampagneLaboratory/gobyweb-plugins. Plugin configuration files offer a version number for each plugin that is displayed on the user interface and makes it possible to track changes to plugin software over time.

**Figure 1. Uploading reads into GobyWeb to create a new Sample.** Read files can be uploaded in a variety of file formats. When the checkbox "Create Multiple Samples" is not selected, individual files are concatenated to yield a single independent biological sample. When the box is not checked, multiple samples are created and associated with the meta-data described on the form.

**Figure 2. Consistent alignment of multiple samples.** GobyWeb supports selecting an arbitrary number of samples for alignment. Configuration of the alignments is entered once through the user interface and applied consistently across all the jobs that will be started.

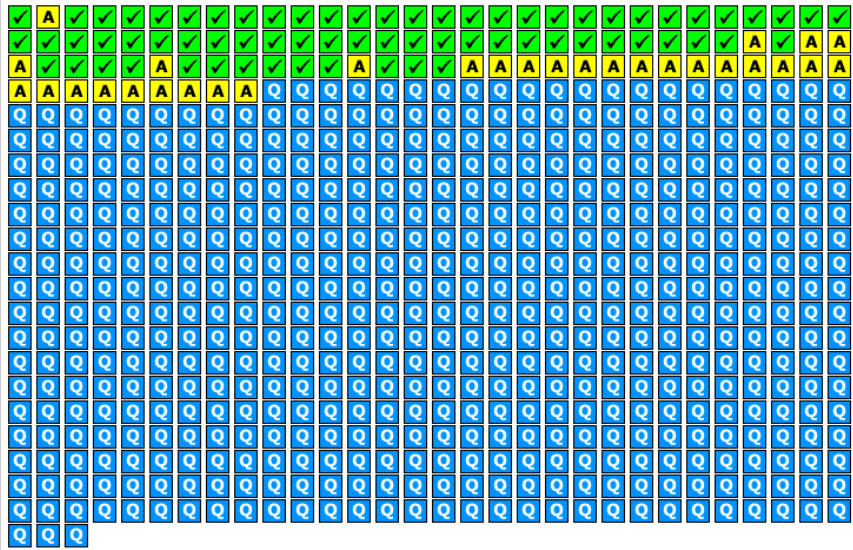

**Figure 3. Visual status for alignment running on compute grid.** The figure shows the visual status for an alignment in progress against a large sample (30GB compressed reads were split into more than 600 chunks and were scheduled for alignment). GobyWeb aligns and sorts each chunk, then concatenates the sorted alignments pieces to yield a completely sorted alignment. Alignments are post-processed to derive base level histograms as well as statistics such as number of aligned reads and number of sequence variations at each cycle.

x

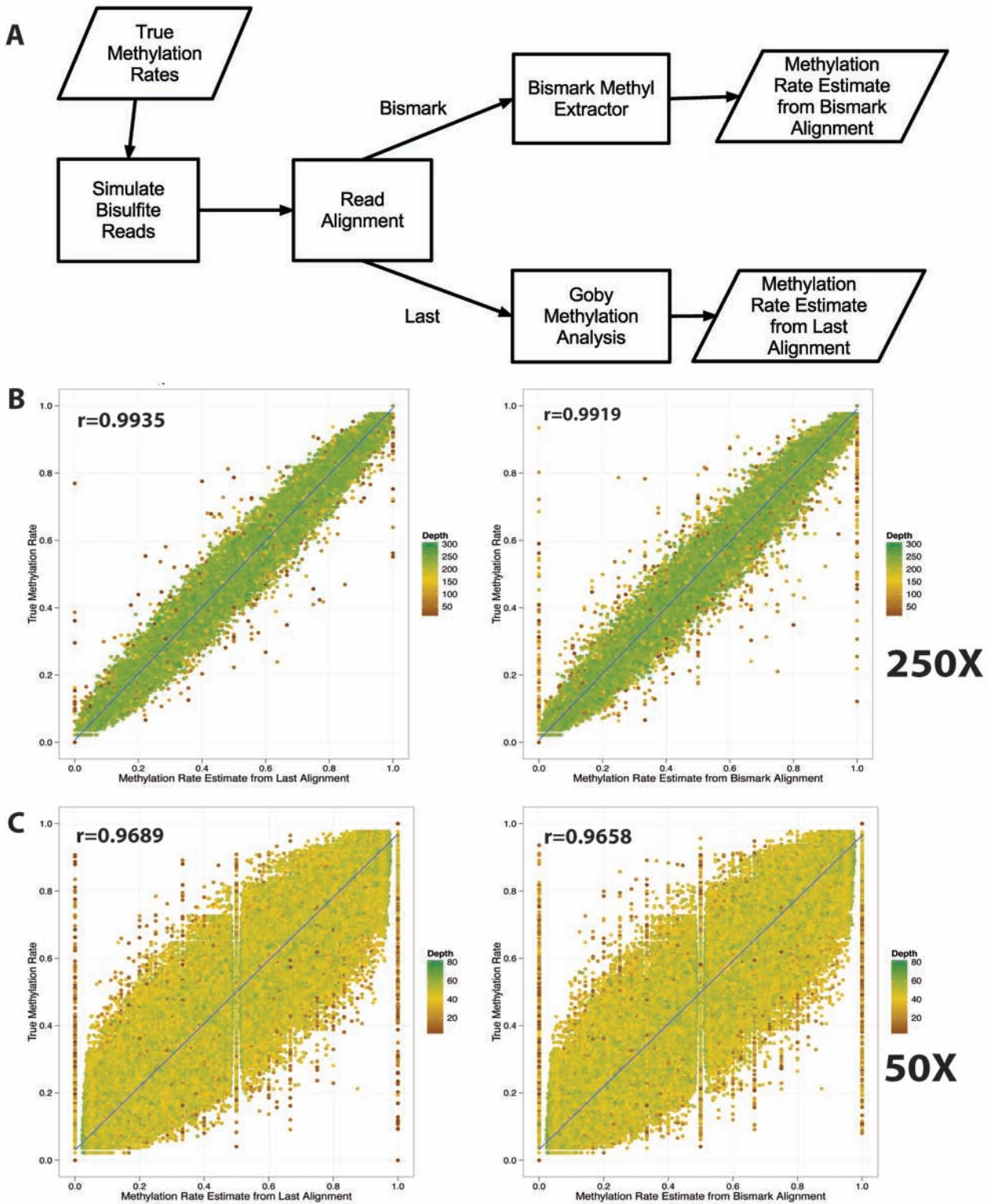

**Figure 4. Comparison between estimates of methylation rates produced with Bismark and Last/Goby.** GobyWeb can align bisulfite converted reads with either the Bismark or the Last aligner. Furthermore, alignments of bisulfite-converted reads can be processed to estimate methylation rates with either Goby or a simple script that post-processes the Bismark result files. Here, (A) we simulated reads from a uniform distribution of methylation rates over a 5MB region of the human genome, at 50X or 250X average coverage and compare the estimate of methylation with the methylation estimate produced by each analysis method. We find (B) that both methods yield comparable agreement with true methylation rates and correlate well with each other when average coverage >50X (data simulated for a target of 50X coverage includes regions of the genome where actual coverage is lower than 50X, these sites tend to have larger disagreement with true methylation).